\documentclass[aps,prd,reprint,showpacs,nofootinbib,superscript address]{revtex4-2}
\usepackage[utf8]{inputenc}
\usepackage{scalerel,stackengine}
\usepackage{physics}
\usepackage{tocbasic}
\usepackage{amssymb}
\usepackage{upgreek}
\usepackage{hhline}
\usepackage{amsmath}
\usepackage{mathtools}
\usepackage[dvipsnames]{xcolor}
\usepackage{multirow,tabularx}
\usepackage{siunitx}
\usepackage{multirow}
\usepackage{graphicx}
\usepackage{etoolbox}
\usepackage{notoccite}
\usepackage{natbib}
\usepackage{mathrsfs}
\usepackage{tensor}
\usepackage{subcaption} % in preamble
\usepackage{accents}
\usepackage{fancyvrb} % in preamble

\def\nn{\nonumber}

\def\be{\begin{equation}}
\def\ee{\end{equation}}
\def\ben{\begin{displaymath}}
\def\een{\end{displaymath}}
\def\bea{\begin{eqnarray}}
\def\eea{\end{eqnarray}}

\DeclareTOCStyleEntry[numwidth=20pt,linefill=\bfseries\TOCLineLeaderFill]{tocline}{section}
\DeclareTOCStyleEntry[entryformat=\textit,numwidth=10pt,linefill=\TOCLineLeaderFill]{tocline}{subsection}
\DeclareTOCStyleEntry[entryformat=\textit,numwidth=10pt,linefill=\TOCLineLeaderFill]{tocline}{subsubsection}

\usepackage{hyperref}
\hypersetup{%
     colorlinks = true,%
     linkcolor = Blue,%
     citecolor = Blue,%
     filecolor = Blue,%
     urlcolor = Blue% 
     }%
\usepackage[capitalize,nameinlink]{cleveref}
\crefname{paragraph}{paragraph}{paragraphs}
\Crefname{paragraph}{Paragraph}{Paragraphs}

\begin{document}

\title{\texttt{Kummitus}: a light-weight toolbox for counting DOF in perturbative QFT}

\author{Carlo Marzo}
\email{carlo.marzo@kbfi.ee}
\affiliation{Laboratory for High Energy and Computational Physics, NICPB, R\"{a}vala 10, Tallinn 10143, Estonia}

\begin{abstract}
The consistent construction of quantum field theories beyond the simplest cases requires a precise characterization of the propagating degrees of freedom. These are encoded in the single-pole structure of two-point functions, a connection firmly established through foundational theoretical work of the last century. While high-level programs for spectral analysis are publicly available, we felt the need to complement the existing landscape with a tool that reaches the gauge-invariant propagator by the shortest possible algorithmic path. This is the purpose of \texttt{Kummitus}, an open-source Wolfram Mathematica toolbox designed to do precisely that: compute the (gauge-invariant) propagator. Beyond its utility in research applications, \texttt{Kummitus} is intended as an accessible and transparent resource for the theoretical community, with particular value for pedagogical purposes.
\end{abstract}
\maketitle

%\tableofcontents

\section{Introduction}
Once the Hilbert space structure is assumed for the description of a quantum-mechanical system, the further imposition of relativistic invariance introduces a natural classification of states in terms of irreducible \emph{unitary} representation of the Lorentz Algebra. Such representations have been completely classified by Wigner and would appear to make any search for the particle content of a quantum theory a dull exercise, solved before its formulation by a preemptive decision on which representations to use. 
Yet, a direct use of canonical operators realizing an explicit unitary form of relativistic invariance (with its fundamental manipulation of the system as space-time translations) is hardly a successful route to account for other complementary facets of Nature. The introduction of non-trivial interactions demands cooperation of other crucial ingredients, embedded in the intertwined requirements of cluster-decomposition, locality and micro-causality.
The current approach\footnote{As any human-made approach to describe reality, it appears inevitable and irreplaceable until it is not. While field-models, with their baggage of diagrammatics, have brought undeniable success, they might be replaced by simpler and less ambiguous methods. Amplitude-first computations are one of these promising alternatives.}, supported by a tremendous predictive success, is to let go of manifest \emph{unitarity} for manifest \emph{locality}. Then, the overabundant particle content introduced by local fields (which forms \emph{non-unitary} representation of the Lorentz Group, containing, at once, multiple irreducible sub-components) unavoidably requires contradictory demands on the parameter space. Hence, a trimming procedure must follow, so to remove pathological states via the imposition of gauge constraints. 
It is a blessing in disguise that particles and local fields are such a poor match as, from the tension coming from their competing requests, only a few of consistent models survive. Much of our trust in the methods of QFT is owned to the phenomenological reach of this surviving bunch. For rank-1 fields, consistency presents QED and Yang-Mills theories. For rank-2, the universal coupling to the energy-momentum tensor ensues. 
It is precisely because of this field-first policy that methods to control and profile the particle content remain under continuous attention from the community \cite{Barker:2024juc,Heisenberg:2025fxc,Hell:2026blj,Barker:2025qmw,Karananas:2024hoh,Lin:2018awc,Lin:2020phk,Percacci:2020ddy,Marzo:2021esg,Mendonca:2019gco}. While it is, in principle, possible to carry out a full spectrography of the relevant degrees of freedom within simple field theories, the growth in the number of indices, followed by the exponential proliferation of particle sectors, calls for an algorithmic approach. Since the original studies of rank-2 and rank-3 theories in geometrically inspired frameworks \cite{Neville:1979rb,Sezgin:1979zf,Sezgin:1981xs}, the use of Spin–Parity projection operators (SPO) has greatly facilitated this process by translating quadratic Lagrangians from their convoluted tensorial form into a simpler manipulation of matrix objects, directly connected with the little-group elements carried by the fields. 
As we will see in the next section, even after the notable simplifications induced by the spin–parity projections, obtaining a \emph{direct} access to the actual particle spectrum still presents significant challenges. Consequently, the full computation of the propagator has rarely been pursued explicitly, with most high-rank analyses favoring a more circumstantial and narrow assessment of the spectrum.
In this work we show that a further step toward removing some of the underlying theoretical assumptions can indeed be taken. In particular, a \emph{direct} construction of the propagator, correctly accounting for the critical gauge identities, is within reach. Within this framework, potential misidentifications of the spectrum can then only arise from implementation errors rather than from intrinsic limitations of the algorithm itself.
To facilitate verification and further developments, we have made the algorithm publicly available through the code \texttt{Kummitus}. Hosted on GitHub, it can be freely inspected, reproduced, and extended by the community\footnote{\texttt{https://github.com/CarloO3/Kummitus}}.
\section{poles $=$ dof}
As well known, in a manifestly unitary approach, fields $\phi_i(x)$ are constructed \emph{a posteriori} as local interpolating functions of creation and annihilation operators associated with a predetermined set of single–particle states $|\mathbf p, i\rangle \sim a_i^\dagger(\mathbf p)|0\rangle $ \footnote{
Here $ |\mathbf p, i\rangle $ denotes a vector belonging to a unitary representation of the Lorentz group, whose components are labeled by an opportune internal index 'i'. We will let go of the index-label, from now on, to lighten the notation. It should nevertheless be understood that an internal index is always assumed for the following manipulations to provide any non-trivial content.}. Once this link to the unitary representations has been established, the corresponding Green functions will exhibit a characteristic behavior that reflects this underlying structure. In the simplest case of a free-theory and a 2-point Green function, we end up with the familiar form     
\begin{align} \label{eqFreeP}
&\langle 0|T\{\phi(x)\phi(y)\}|0\rangle
\equiv \nonumber \\
&\Delta_0(x-y) =
\int \frac{d^4p}{(2\pi)^4}
\frac{i\,e^{-ip\cdot(x-y)}}{p^2-m^2+i\epsilon}. 
\end{align}
The connection between the singular momentum behavior and the physical notion of particles is made precise by the LSZ reduction procedure. This mandates that the Green functions are to be multiplied by factors of $p^2 - m^2$ in order to project onto physical asymptotic states. Taking the on-shell limit $p^2 = m^2$ then isolates the contribution of the single-particle pole. The presence of this pole is therefore essential in ensuring that, after the LSZ factors are applied and the on-shell limit is taken, the resulting scattering amplitude remains finite and non-vanishing.
Possibly more directly, Eq.~(\ref{eqFreeP}) is the manifestation of the general, non-perturbative connection between the propagator and the (Källén–Lehmann) spectral function $\rho(M^2)$ 
\begin{align} \label{eqKL}
& \rho(\mu^2)
=
(2\pi)^3
\sum_n
\delta^{(4)}(p - p_n)
\left|
\langle 0 | \phi(0) | n \rangle
\right|^2 , \\
& \hspace{3.4cm} (p_n^2 = \mu^2, p_0 > 0) \nonumber
\end{align}
resulting in the general form 
\begin{align} 
\Delta(x-y)
=
\int \frac{d^4p}{(2\pi)^4}
i\,e^{-ip\cdot(x-y)} \int d \mu^2 \frac{\rho(\mu^2)}{p^2-\mu^2+i\epsilon} \nonumber
\end{align}
and linking the full, complete basis of states $| n \rangle$ to the singular behavior of the propagator in momentum space. 

\section{Fields, ghosts and tachyons}
Model-building of complex, realistic theories proceeds in the opposite direction: locality and Lorentz invariance are automatic, while unitarity must be 'prayed' for. In particular, the typical properties of unitary propagation we have met, as the propagator's large momentum behaviour $\Delta(p) \sim 1/p^{2} $  and the positivity of the spectral function Eq.~(\ref{eqKL}), are not a general consequence and must be recovered. 
Deviations from these two requirements signal a breach of unitarity commonly referred to under the generic nomenclature of \emph{ghost propagation}. They provide, however, two different requirements that our candidate theory must possess. The first, dispose of high-derivative terms in the local quadratic Lagrangian (without the need of any reference to classical Ostrogradsky instabilities). The second, tunes the parameters of the theory so to have only positive residues \footnote{Both statements suffer from small nuances which we explicit here. First, the overall sign of the Lagrangian is irrelevant for scattering computations, so the precise request is that the residues of all particles \emph{have the same sign}. Secondly, as playing with high-derivative field redefinitions reveals, it is possible to have unitary and causal theories with spurious high-derivatives which do not affect the large-momentum behavior of the propagator.}.
Finally, the causal texture also needs to be enforced, given that we have to ensure the creation of physical states with zero or positive masses in order to avoid tachyonic propagation. Again, this will translate into a further requirement for those combination of parameters defining the zeroes of $p^2$.
\section{Projection operators and the saturated propagator}
Spin-parity operators introduce enormous practical simplifications in illuminating the possible particle content of a field-valued Lagrangian. Relevant for our discussion is that they can directly channel the gauge-invariant propagator itself and, therefore, the (perturbative) spectral function. A large amount of theoretical effort has improved and adopted the projection operator approach supporting a steady progress in seeking more realistic proposals of new Physics. We refer the reader to \cite{Percacci:2025oxw,Barker:2025rzd,Barker:2025fgo,Barker:2025qmw,Karananas:2014pxa,Lin:2018awc,Lin:2020phk,Barker:2024ydb}  for some recent applications. In particular, we will use the notation introduced in \cite{Percacci:2020ddy} for all further manipulations and we only present here the minimal definitions necessary in order to follow the subsequent reasoning. 
For the \emph{factory} version of \texttt{Kummitus}, the dynamic generated using tensor fields up to rank-3 can be promptly investigated. Different type of fields can be added by simply updating the necessary set of SPOs. Following the classification of \cite{Percacci:2020ddy}, the irreducible particle sectors contained in the reducible fields are summed up to the intimidating list 
\begin{align}\label{eq1}
A_{\mu\nu\rho} \supset\; & 3_1^- \oplus 2_1^+ \oplus 2_2^+ \oplus 2_3^+ \oplus 2_1^- \oplus 2_2^- \notag \\
&\oplus 1_1^+ \oplus 1_2^+ \oplus 1_3^+ \notag \\
&\oplus 1_1^- \oplus 1_2^- \oplus 1_3^- \oplus 1_4^- \oplus 1_5^- \oplus 1_6^- \notag \\
&\oplus 0_1^+ \oplus 0_2^+ \oplus 0_3^+ \oplus 0_4^+ \oplus 0_1^-, \notag \\
H_{\mu\nu} \supset\;& 2_4^+ \oplus 1_7^- \oplus 0_5^+ \oplus 0_6^+, \notag \\
V_\mu \supset\;& 1_8^- \oplus 0_7^+, \notag \\
S \supset\;& 0_8^+ \, , 
\end{align}
where the self-evident symbolic notation $\text{Spin}^{\text{Parity}}$ is adorned with a subscript to distinguish different occurrences of the same spin/parity sector.
We can use the compact notation
 \[
\Phi = \left\lbrace  S, V_{\mu}, H_{\mu \nu},  A_{\mu \nu \rho} \cdots\right\rbrace ,
\]
to represent the set of all fields, of possibly different rank, that mix at the quadratic level. If such a mixing is not present, the (perturbative) spectral problem can be separated into independent branches. The corresponding quadratic Action can therefore be put in the generic form (formulas and notation taken from \cite{Marzo:2021esg}, to which we refer for details)
{\small
\begin{equation}\label{eqQL}
\mathcal S =  \frac{1}{2} \int d^4 q \, \bigg( \Phi(-q) \,  \mathcal \, K(q) \, \Phi(q) + \mathcal{J}(-q)\Phi(q) + \mathcal{J}(q)\Phi(-q) \bigg)\, . 
\end{equation}
}
%$P^{i,j}_{\left\lbrace S,p\right\rbrace}$ removed from next line
In Eq.~(\ref{eqQL}) a classical \emph{Schwinger} source is introduced as a proxy for interactions. The SPO {\small $P^{i,j}_{\left\lbrace S,p\right\rbrace}$} possess a rather involved index structure, reflecting their role in connecting the $i$-th representation of spin $S$ and parity $p$ to the corresponding $j$-th representation with the same spin and parity. They form a complete set of orthogonal and hermitian operators that satisfy the algebra 
\bea \label{eq6}
&&\sum_{S,p,i} P^{i,i}_{\left\lbrace S,p\right\rbrace}{}_{\mu_1 \mu_2 \cdots \mu_n}^{\quad\quad \nu_1 \nu_2 \cdots \nu_n} = \displaystyle{\hat 1}{}_{\mu_1 \mu_2 \cdots \mu_n}^{\quad\quad \nu_1 \nu_2 \cdots \nu_n} , \nn \\ &&\nn \\
&& P^{i,k}_{\left\lbrace S,p\right\rbrace}{}_{\mu_1 \mu_2 \cdots \mu_n}^{\quad\quad \rho_1 \rho_2 \cdots \rho_n} \,\, P^{j,w}_{\left\lbrace R,m\right\rbrace}{}_{\rho_1 \rho_2 \cdots \rho_n}^{\quad\quad \nu_1 \nu_2 \cdots \nu_n} = \nn \\
&& \delta_{k, j}\, \delta_{S, R}\, \delta_{p, m}\, P^{i,w}_{\left\lbrace S,p\right\rbrace}{}_{\mu_1 \mu_2 \cdots \mu_n}^{\quad\quad \nu_1 \nu_2 \cdots \nu_n} , \nn \\ &&\nn \\
&& P^{i,j}_{\left\lbrace S,p\right\rbrace}{}_{\mu_1 \mu_2 \cdots \mu_n}^{\quad\quad \nu_1 \nu_2 \cdots \nu_n} = 
\left(P^{j,i}_{\left\lbrace S,p\right\rbrace}{}^{\nu_1 \nu_2 \cdots \nu_n}_{\quad \quad \mu_1 \mu_2 \cdots \mu_n}\right)^*  . \nn \\
\eea
By enforcing the identity 
\bea \label{eq4}
\phi_{\mu_1 \mu_2 \cdots \mu_n}\left(q\right) = \sum_{S,p,i} P^{i,i}_{\left\lbrace S,p\right\rbrace}{}_{\mu_1 \mu_2 \cdots \mu_n}^{\quad\quad \nu_1 \nu_2 \cdots \nu_n}\left(q\right) \phi_{\nu_1 \nu_2 \cdots \nu_n}\left(q\right), \nn 
\eea
we finally can remove the tangled layer of Lorentz indices and highlight the particle sectors contained in the kinetic term
{\small
\begin{eqnarray}\label{eq8}
&&\int d^4 q \,  \Phi(-q) \, K(q) \, \Phi(q) \, =  \nn \\
&& = \int d^4 q \, \Phi(-q) \, \sum_{S,p,i,j} \bigg( a^{\left\lbrace S,p \right\rbrace}_{i,j}  P^{i,j}_{\left\lbrace S,p \right\rbrace} \bigg) \, \Phi(q)  \,\,. \nn
\end{eqnarray}
}
The final product is a complete rewriting of the Lagrangian in terms of the matrices $a^{\left\lbrace S,p \right\rbrace}_{i,j}$ which uniquely define the theory, while eliminating the complications introduced by high-rank fields. This dual description provides an alternative representation for the main objects that commonly appear in conventional field-theoretic computations. In particular, the computation of the propagator $\mathcal D(q)$ proceeds through the equation
\bea \label{eq10}
&& K(q) \, \mathcal D(q) \, =   \sum_{S,p,i,j} \bigg( a^{\left\lbrace S,p \right\rbrace}_{i,j}  P^{i,j}_{\left\lbrace S,p \right\rbrace} \bigg) \, \mathcal D(q) =  \hat{1} \,\,, \nn
\eea
which admits the solution 
\bea \label{eq10b}
\mathcal D(q) = \sum_{S,p,i,j}  b^{\left\lbrace S,p \right\rbrace}_{i,j}  P^{i,j}_{\left\lbrace S,p \right\rbrace} \,\,, \nn
\eea 
with $b^{\left\lbrace S,p \right\rbrace}_{i,j} = \left(a^{\left\lbrace S,p \right\rbrace}_{i,j}\right)^{-1}$: a simple matrix inversion problem. 
Even simpler is the profiling of gauge symmetries, which appear in the degeneracy of the $a^{\left\lbrace S,p \right\rbrace}_{i,j}$ matrices and are in direct correspondence with their null-vectors $X_i^{s}$. For each of them, the action will be left unchanged under the transformations
\bea \label{eq11}
\delta \Phi = X^s_i \, P^{i,j}_{\left\lbrace S,p \right\rbrace} \Psi \,, \quad\,\,(s = 1,2\cdots n)\,.
\eea
This constructive definition of gauge symmetries has allowed a full scan of healthy models in \cite{Barker:2025fgo,Barker:2025rzd}.
The critical fixing of the gauge, which is required in order to perform perturbative computations, must then be implemented in the SPO formalism in order to access the theory spectrum. This can be achieved by arbitrarily selecting the non-degenerate sub-matrix $\tilde{a}^{\left\lbrace S,p \right\rbrace}_{i,j}$. The freedom in this choice is in direct correspondence with the gauge-parameter dependence encountered in the standard gauge-fixing procedure. It is, ultimately, completely balanced by a parallel imposition of \emph{gauge constraints} on the sources
\bea \label{eq12}
X^{*}{}^s_j \,\, P^{i,j}_{\left\lbrace S,p \right\rbrace} \,\, J \left(q\right) = 0 \,, \quad\,\,(s = 1,2\cdots n)\,.
\eea
The particular solution of Eq.(\ref{eq12}), $\tilde{J}(q)$, is the last obstacle for the definition of the theory's gauge-invariant \emph{saturated} propagator 
\bea \label{eq13}
\mathcal D_S(q) =  \tilde{J}^*\left(q\right) \bigg(\sum_{S,p,i,j}  \tilde{b}^{\left\lbrace S,p \right\rbrace}_{i,j}  P^{i,j}_{\left\lbrace S,p \right\rbrace}\bigg)  \tilde{J}\left(q\right) \, .
\eea
As illustrated in the previous section, and further supported by an absence of spurious degrees of freedom connected to the gauge-fixing procedure, we can access the \emph{physical} spectrum by analyzing the isolated poles of $\mathcal D_S(q)$.

\section{\texttt{Kummitus}: raison d'être}
While the previous definitions are straightforward, the explicit computation of $\mathcal D_S(q)$ is impeded by a number of intermediate obstacles of varying difficulty. Some of these, such as the transition to momentum space, the computation of the matrices $a^{\left\lbrace S,p \right\rbrace}_{i,j}$, and the characterization of the null vectors generating the gauge symmetries, can be handled by a somewhat direct call of the linear-algebra tools of Wolfram Mathematica \cite{Mathematica}, and the tensorial routines provided by \texttt{xAct} \footnote{\texttt{http://www.xact.es/}}\cite{Nutma:2013zea,Brizuela:2008ra}.
An harder challenge is posed by the computation of the gauge constraints Eq.(\ref{eq12}), which involve (many) tensorial equations which become untreatable for all but the simplest cases. \\
These obstacles have favoured ways to assess the spectrum reliant on \emph{indirect} methods (a notable exception is found in \cite{Lin:2018awc,Lin:2020phk} ). For instance, completely skipping the computation of the gauge constraints, we can assume that all the poles $q^2 = M_n^2$ of the propagator are zeroes of the determinant of any of $a^{\left\lbrace S,p \right\rbrace}_{i,j}$ and infer their residue from the computation of the trace of the (gauge-fixed) matrix  
$\tilde b^{\left\lbrace S,p \right\rbrace}_{i,j} = \left(\tilde a^{\left\lbrace S,p \right\rbrace}_{i,j}\right)^{-1}$. This procedure has supported an algorithm to evaluate the spectrum which make use of (variations) of the expression 
\begin{align} \label{shortcut}
\underset{q^2 \rightarrow M_n^2}{{\bf Res}}\quad\,\sum_i\, \left(-1\right)^p\,\tilde{b}^{\left\lbrace S,p \right\rbrace}_{i,i} \,> 0 \,.    
\end{align}
This common approach falls short in situations involving massless particles, degenerate spectra, or collective gauge-invariant descriptions of massive modes. A case in point is the massless Fierz-Pauli spin-2 model, which lies beyond the reach of Eq.~(\ref{shortcut}): the latter would miss the interplay between the spurious poles in the $2^+$ and $0^+$ sectors, whose cancellation is precisely what ensures the propagation of a single helicity-2 state.
Such critical, highly symmetrical models involving the cooperation of multiple components, far from representing an exotic branch of research, are a most promising start for a non-linear completion \cite{Henneaux:1989jq,Barnich:2025jna,Barnich:2000zw,Henneaux:1997bm,Barnich:2017nty,Berends:1984rq,Fang:1978rc,Barnich:1993vg,Deser:1963zzc,Barker:2025fgo,Barker:2025rzd,Marzo:2021iok,Marzo:2024pyn} , which is, ultimately, the goal of profiling non-pathological free propagation. 
%
%Our purpose is to offer a minimal computational tool that can directly account for the full structure of the propagator without relying on unnecessary simplifications.   
%
\section{\texttt{Kummitus}: Method}
The \texttt{Kummitus} toolbox is designed to address the propagator directly, and its algorithm is accordingly reduced to its bare essentials, following a straightforward implementation of the definitions of \cite{Percacci:2020ddy}. Given a set of fields, the analysis requires as a prerequisite the corresponding set of SPOs. The first release of \texttt{Kummitus} includes built-in SPOs for fields up to rank 3 \cite{Mendonca:2019gco,Percacci:2020ddy}. For higher-rank cases \cite{Bittencourt:2025roa}, the algorithm for generating the required SPOs is systematic, and their inclusion, if needed, can be easily carried out. All subsequent steps leading to the matrices $a^{\left\lbrace S,p \right\rbrace}_{i,j}$ follow directly from their definitions in Ref.~\cite{Percacci:2020ddy}, implemented via the tensorial routines of \texttt{xAct} and the linear-algebra tools of Wolfram Mathematica.

More careful design is required for the correct and complete treatment of the gauge constraints and the subsequent inversion problem leading to the fully saturated propagator. As mentioned, this is often, though not always~\cite{Lin:2018awc,Lin:2020phk}, the point at which spectral analyses stall, with the spectrum assessed through indirect arguments rather than direct computation. To overcome the challenge posed by the tensorial system Eq.~(\ref{eq12}), we exploit the covariance of the problem by working in a suitably chosen reference frame.
The candidate poles are identified as the zeros of the non-degenerate matrices $\tilde{a}^{\left\lbrace S,p \right\rbrace}_{i,j}$. For candidate massive poles, $q^2 = M_n^2$, the tensors are expanded in components in the rest frame $q = (\omega, \vec{0})$, reducing Eq.~(\ref{eq12}) to an algebraic system admitting a straightforward solution. For massless poles, $q^2 = 0$, the tensors are instead expanded in the basis $q = (\omega, 0, 0, \kappa)$, with the light-like limit $\kappa \rightarrow \omega$ deferred until the propagator has been fully computed.

In either frame, the solutions of Eq.~(\ref{eq12}) take a simple form in terms of an independent subset $j_i$ of the original unconstrained components. This generalizes familiar results: the paradigmatic case is the Ward identity derived from abelian gauge invariance,
\begin{align}
   q_{\mu} J^{\mu} = 0 \;\to\; \omega J^{0} = \kappa J^{4}\,,
\end{align}
which enforces the mutual cancellation between timelike and longitudinal components. The Wolfram Mathematica series-expansion routines are then employed to compute the on-shell singular limits, organizing the saturated propagator as a Laurent series of the form
\begin{align} \label{eqxx}
\lim_{q^2 \to M_n^2} D_{S}(\omega) \sim \frac{J_i^\dagger \,\mathcal{K}_{ij}\, J_j}{\left(q^{2}-M_n^{2}\right)^r} + \cdots
\end{align}
The pole order is decisive: $r > 1$ signals the presence of ghosts, while $r \neq 1$ exposes the candidate pole as spurious, despite having appeared among the zeros of the determinants. For simple poles, $r = 1$, the residue must be extracted to determine both the number of propagating degrees of freedom and the sign of the norm of the corresponding single-particle states. This is achieved by diagonalizing the numerator in Eq.~(\ref{eqxx}), reducing the propagator to the diagonal form
\begin{align} \label{eqyy}
\lim_{q^2 \to M_n^2} D_{S}(\omega) \sim \frac{\displaystyle\sum^{\text{states}}_{m} r_m\, |j^m|^2}{\left(q^{2}-M_n^{2}\right)}\,,
\end{align}
where the eigenvalues $r_m$ and eigenvectors $j^m$
of the numerator matrix complete the identification of the propagating degrees of freedom and their physical norm.
\section{AI assistance and spurious poles}
The idea behind \texttt{Kummitus} is to provide the community with a tool that is minimal in 'just compute the propagator'. It grew out of a collection of user-\emph{unfriendly} routines that the author\footnote{The idiosyncrasies of the code reflect the fact that \texttt{Kummitus} was written by a theoretical physicist rather than a software developer, coding being, for the author, a means to a theoretical end rather than a discipline. By making the code publicly available, users (humans and not) are invited to expand and improve upon the core implementation at will.} used for many years, assembled and painstakingly tuned for particular applications, rather than built for general use.
A central technical difficulty stemmed from a naive reliance on Wolfram's \texttt{Solve} function, which may select pivot variables whose expressions develop spurious divergences in the massless on-shell limit $\kappa \to \omega$ ($q^2 \to 0$). This can lead to the erroneous flagging of otherwise healthy theories as pathological~\footnote{That the original implementation was at risk of incorrectly flagging healthy theories due to spurious poles in the solution of the source constraints was brought to the author's attention by Will Barker.}. The practical workaround involved a direct inspection of the solutions and, in an attempt to automate the full workflow, a systematic scan over candidate subsets of the source components until a regular solution was identified.

The recent advances in large language models (LLMs) for software development provided an opportunity to replace this provisional workaround with a more robust solution. The result is the \texttt{SystemSolver} module, which supersedes the direct call to \texttt{Solve}, and whose integration into the main code marks a transition of \texttt{Kummitus} into a form ready for public release. The risk of leaning too heavily on artificial assistance is, of course, that one ends up with black boxes of obscure functionality, forced to rely on \emph{faith} rather than understanding, in assessing the robustness of the output. This concern is addressed here in two ways. First, the \texttt{SystemSolver} module is built through incremental, well-understood improvements over the original human-written routines. Second, the problem itself admits a natural consistency check: the source solutions can be back-substituted directly into the system of equations, therefore providing a strong verification of correctness. 
We used the assistance of Claude Code \cite{claudecode2024} for the public version of \texttt{Kummitus}.

\section{Usage and examples}
We now illustrate the basic workflow required to access the saturated propagator and the associated spectral properties initiated by a call to \texttt{Kummitus.wl}. We do so by walking through a selection of explicit examples drawn from the accompanying notebook \texttt{Models.nb}, which covers a range of applications from the simplest cases to more recent higher-spin models\footnote{The formatting of big formulas is error-prone. We refer the reader to the notebook for the full digitalization of the presented Lagrangians.}. 
\subsection{Helicity 1: Maxwell Theory}
\begin{figure}[h]
  %  \centering
    \includegraphics[width=0.8\linewidth]{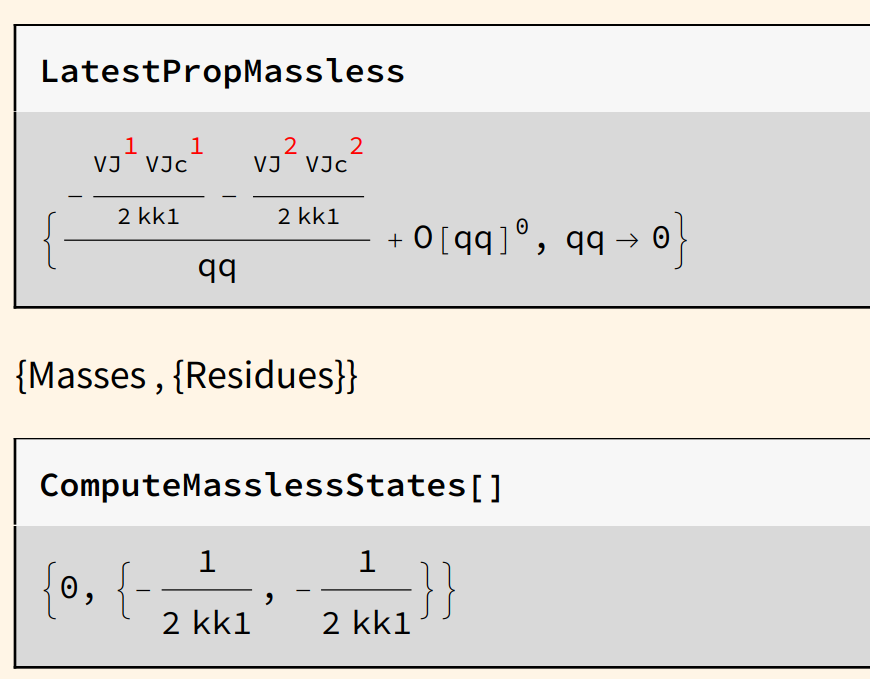}
   \caption{Saturated propagator and (two) residues for the Maxwell theory. Output from the provided notebook \texttt{Models.nb}.}
    \label{fig:MaxWellFinal}
\end{figure}

The user has only one input to provide: the quadratic Lagrangian in terms of any of the four fields \texttt{S[]}, \texttt{V[m]}, \texttt{H[m,n]} and \texttt{A[m,n,r]} and the derivative operator \texttt{dd[m][field]}. Up to here, these are simple proxies for analogue definitions taken from \texttt{xAct}. For a study of the Maxwell model, this entails a simple input
{\small
\begin{verbatim}
    model = kk1 (dd[r]@V[s] dd[-r]@V[-s] 
    - dd[-s]@V[s] dd[-r]@V[r]) ;
\end{verbatim}
}
At the core of \texttt{Kummitus} lies the reduction of tensorial expressions to their component form. For high-rank fields, this step can be computationally expensive, however, since it is required only once per field content, it has been decoupled from the spectral analysis itself. The user invokes \texttt{SourceDeclare[model]}, which identifies the field content of the model and generates the tensor-to-component reduction rules. Once declared, any number of Lagrangians sharing the same field content can be analysed without repeating this step. Should new fields be introduced, a second call to \texttt{SourceDeclare} is recommended to update the reduction rules. This is not strictly necessary, the system will automatically detect any undeclared sources and handle them accordingly, but doing so avoids redundant recomputation and improves efficiency for repeated calls. 
The computation proceeds by invoking the main function \texttt{CompSatProp[model]}, which implements the algorithm described above and returns the saturated propagator in the form of Eq.~(\ref{eqyy}). Throughout execution, the printed output reports the main steps of the computation explicitly. 
It begins by identifying which sectors from the list Eq.~(\ref{eq1}) can propagate. The four slots $\{\cdot,\cdot,\cdot,\cdot\}$ indicate the field to which each sector belongs, with the leftmost entry corresponding to the rank-3 tensor and the rightmost to the scalar. In this simple case, for instance, the sector $1^-$ admits the single entry $1^-_8$ and similarly the $0^+$ sector admits $0^+_7$:
\begin{verbatim}
=== Current Field Configuration ===

Fields: {vector}

J^P Sectors: {{1,m},{0,p}}

Representation Tables:

  sector[1,m] = {{},{},{8},{}}

  sector[0,p] = {{},{},{7},{}}

=== Field Content Configuration ===

Detected fields: {vector}
\end{verbatim}
The computation proceeds then, by accounting for the gauge constraints 
\begin{Verbatim}[commandchars=\\\{\}]
Total source constraints (vincoli): 1

=== Source Constraints ===

vincoli[[1]] =  \( \frac{\text{q\textsubscript{m} q\textsubscript{n} VJ\textsuperscript{n}}}{qq} \)
\end{Verbatim}
which clearly shows the known transversality requirement (\texttt{qq} being the symbol for $q^2$, the squared four-momentum) $q\cdot J = 0$.
This is solved, as discussed, in component space. The solution is then shown in the last printed lines 
\begin{Verbatim}[commandchars=\\\{\}]
=== Solution ===
--- Branch 1 ---
  Independent: \(  \{\text{VJ\textsuperscript{{\color{red}0}},VJ\textsuperscript{{\color{red}1}},VJ\textsuperscript{{\color{red}2}}}\}  \)
  \( \text{VJ\textsuperscript{{\color{red}3}}}\to \frac{\omega \text{VJ\textsuperscript{{\color{red}0}}}}{\kappa} \)
\end{Verbatim}
Clearly, after sending $\kappa \to \omega$ ($q^2 \to 0$), this reproduces the known identity between the longitudinal and temporal components of the vector source. After a successful call to \texttt{CompSatProp} we can appreciate the explicit form of the saturated propagator by a call to \texttt{LatestPropMassless} which outputs the Laurent expansion around the (massless) pole FIG.~(\ref{fig:MaxWellFinal}), highlighting the expected two-helicity propagation
Although an overkill for this simple example, the discussed diagonalization can be carried on by the \texttt{ComputeMasslessStates[]} routine, which outputs the values of the pole residues (stored in the global variable \texttt{masslessData}).

\subsection{Massless spin-2: Einstein theory and helicity decomposition}
\begin{figure}[h]
    \centering
    \includegraphics[width=1.05\linewidth]{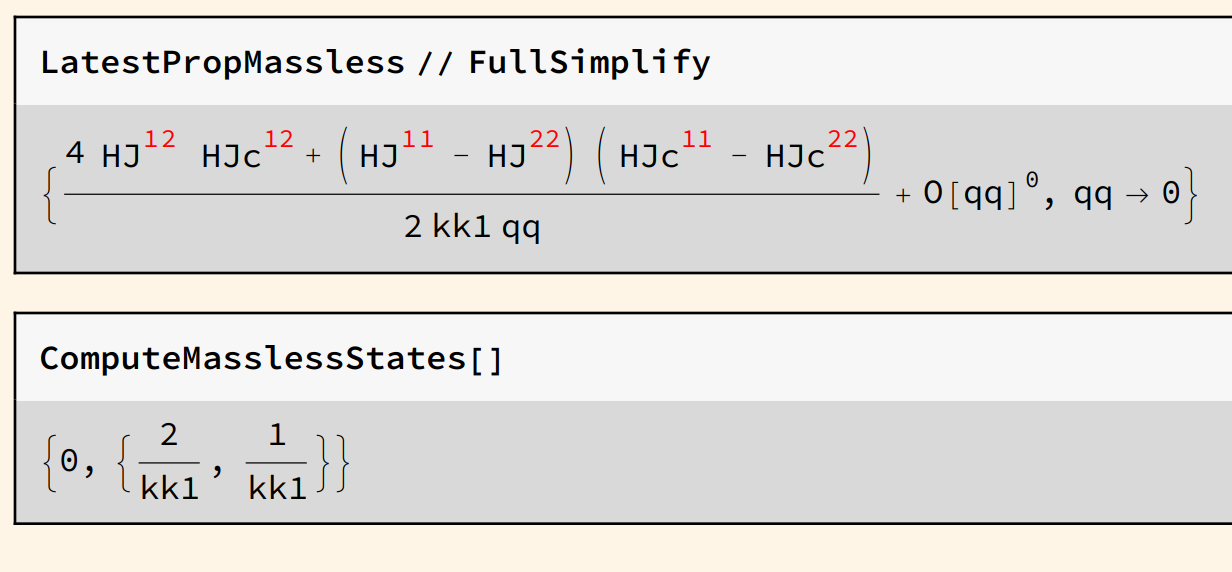}
    \caption{Saturated propagator and residues for Einstein quadratic Lagrangian}
    \label{fig:EinsteinProp}
\end{figure}

\begin{figure}[h]
    \centering
    \includegraphics[width=1.\linewidth]{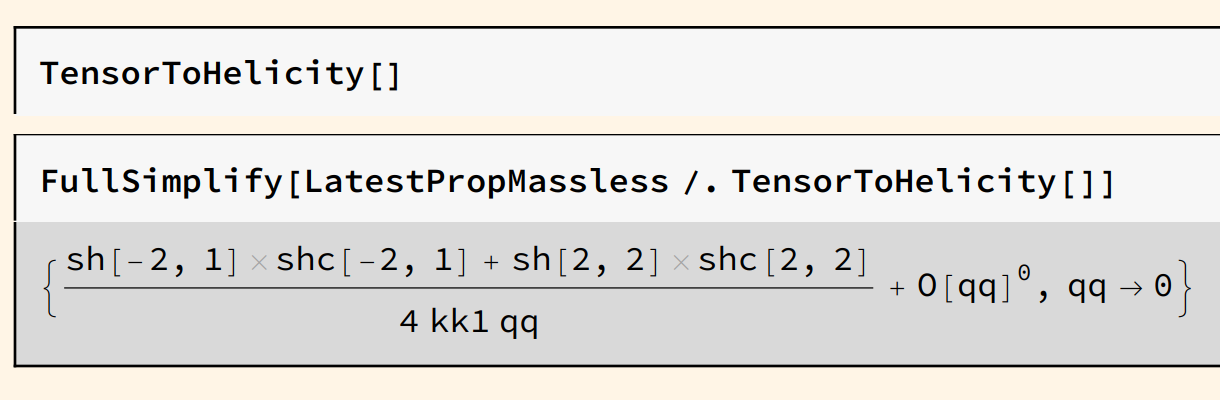}
    \caption{Saturated propagator for Einstein theory in helicity basis.}
    \label{fig:EinsteinPropHel}
\end{figure}

We can readily tackle a theory of massless spin-2 as obtained by imposing linear diffeomorphism 
\begin{align} \label{lindiff}
    \delta H_{mn} = \partial_m \xi_n(x) + \partial_n \xi_m(x)\, , 
\end{align}
to the generic quadratic (two-derivative) local lagrangian of a rank-2 symmetric field. 
The invariant result is:
\begin{align} \label{einst}
\mathcal{L} = & \kappa_1 \Big(
 H^{ab}\partial_c \partial_b H_a{}^c - H^{ab}\partial_b \partial_a H^c{}_c
 \notag \\
&+ \frac{1}{2} H^a{}_a \Box H^b{}_b
- \frac{1}{2} H^{ab}\Box H_{ab}\Big) \, .
\notag \\ 
\end{align}
Repeating the same chain of calls to \texttt{Kummitus} we are immediately presented with a more populated particle sector
\begin{verbatim}
=== Current Field Configuration ===
Fields: {rank2sym}

J^P Sectors: {{2,p},{1,m},{0,p}}

Representation Tables:
  sector[2,p] = {{},{4},{},{}}
  sector[1,m] = {{},{7},{},{}}
  sector[0,p] = {{},{5,6},{},{}}
\end{verbatim}
and the corresponding reductions due to (linear) diffeomorphism 
\begin{Verbatim}[commandchars=\\\{\}]
=== Solution ===
--- Branch 1 ---
  Independent: \(  \{\text{HJ\textsuperscript{{\color{red}00}},HJ\textsuperscript{{\color{red}01}},HJ\textsuperscript{{\color{red}02}},HJ\textsuperscript{{\color{red}11}},HJ\textsuperscript{{\color{red}12}},HJ\textsuperscript{{\color{red}22}}}\}  \)
  
  \( \text{HJ\textsuperscript{{\color{red}03}}}\to \frac{\omega \text{HJ\textsuperscript{{\color{red}00}}}}{\kappa} \)
  \( \text{HJ\textsuperscript{{\color{red}13}}}\to \frac{\omega \text{HJ\textsuperscript{{\color{red}01}}}}{\kappa} \)
  \( \text{HJ\textsuperscript{{\color{red}23}}}\to \frac{\omega \text{HJ\textsuperscript{{\color{red}02}}}}{\kappa} \)
  \( \text{HJ\textsuperscript{{\color{red}33}}}\to \frac{\omega\textsuperscript{2} \text{HJ\textsuperscript{{\color{red}00}}}}{\kappa\textsuperscript{2}} \)
\end{Verbatim}
We can again have a direct look at the propagator FIG.(\ref{fig:EinsteinProp}) to reveal,
as expected, that a healthy propagation of two independent states is possible. This is an opportune moment to pause and consider (while unnecessary for the quantitative counting of dofs, which is performed in a basis-independent way in \texttt{ComputeMasslessStates[]} and \texttt{ComputeMassiveStates[]}) the particular source-dependent form of the saturated propagator visible after calling \texttt{LatestPropMassless}. When formulated in terms of the sources \texttt{HJ[a,b]}, the nature of the propagated helicity is not explicit. In this simple case it is easy to track which sector is propagating, by directly looking at where the pole is originated. But for more involved scenarios with simultaneous multisector propagation, such information is key. To reveal it, we rewrite the source components in terms of helicity eigenstates (as already done in the previous works \cite{Barker:2025fgo,Barker:2025rzd}). 
The alternative form becomes the more elegant FIG.(\ref{fig:EinsteinPropHel})
where the objects \texttt{sh} and \texttt{shc} replace the sources \texttt{HJ} and \texttt{HJc}. 
The symbol \texttt{sh} stands for the helicity amplitude with helicity eigenvalues given by the first slot (which is $\pm 2$ in FIG.(\ref{fig:EinsteinPropHel})). The second slot stands for an extra index to list the degeneracy within a given helicity eigenvalue. 

\subsection{TDIFF is enough}
As known \cite{vanderBij:1981ym,Alvarez:2006uu,Carballo-Rubio:2022ofy,Percacci:2017fsy}, a smaller symmetry than Eq.~(\ref{lindiff}) is enough to describe healthy propagation of a massless spin-2 particle. This is achieved restricting the vector generator of linear diffeomorphism Eq.~(\ref{lindiff}) to be transversal $\partial_m \xi^m(x) = 0$. The resulting Lagrangian presents a slightly more involved structure:
\begin{align}
&\mathcal{L} = \;
k_1 \Bigl[ H^{ab}\partial_b \partial_a H^c{}_c \Bigr] \notag + k_2 \Bigl[ H^{ab}\partial_c \partial_b H_a{}^c
- \frac{1}{2} H^{ab}\Box H_{ab} \Bigr] \notag \\[6pt]
&+ k_3 \Bigl[ H^a{}_a\,\partial_c \partial_b H^{bc} \Bigr]+ k_4 \Bigl[ m_1^2 H^a{}_a H^b{}_b
+ H^a{}_a\,\Box H^b{}_b \Bigr] \, .
\end{align}
After feeding it to \texttt{Kummitus}, 
\begin{figure}[h]
    %\centering
\includegraphics[width=1.05\linewidth]{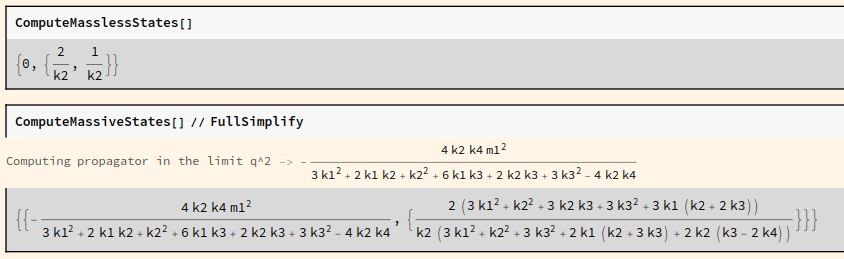}
    \caption{Massless and massive propagation under linear transversal diffeomorphism.}
    \label{fig:TDIFF}
\end{figure}
the additional presence of a massive scalar is promptly detected FIG.~(\ref{fig:TDIFF}). While cumbersome (due to our choice of an 'unbiased' parameterization of the Lagrangian), a simple demand of positivity for all mass and residues reveals a unitary setup. 

\subsection{Massive spin-2: Fierz-Pauli}
\begin{figure}[h]
    \centering
    \includegraphics[width=0.9\linewidth]{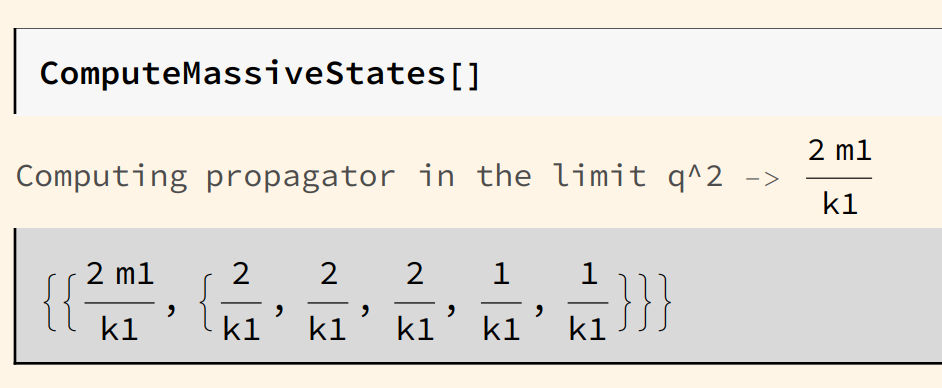}
    \caption{Masses and the five residues for the propagator of Fierz-Pauli theory from \texttt{Kummitus}}
    \label{fig:FierzPauli}
\end{figure}

An addition to Eq.(\ref{einst}) of the mass term
\begin{align} \label{FierzPauliMass}
    \mathcal{L}_m = m_1 (H_a{}^a H_b{}^b - H_{ab} H^{ab})
\end{align}
allows us a study of a gauge-less case for massive spin-2 propagation. The absence of gauge constraints is also pointed out 
\begin{verbatim}
Converting constraints to rest frame...
Constraints in rest frame: 0 equations
Solving constraints...
No constraints - all variables are free.
\end{verbatim}
as it is the position of the pole
\begin{verbatim}
=== Mass Spectrum ===
Unique poles (qq = m^2):
  m^2 = (2 m1)/k1
\end{verbatim}
Less illuminating is the shape of the full saturated propagator in the singular (massive) limit obtained by calling \texttt{LatestPropMassive}. Nevertheless, the basis-independent diagonalization clarifies the number of independent residues and the corresponding parametric dependence FIG.(\ref{fig:FierzPauli})

\subsection{Gauge-symmetric massive spin-2 - Part 1}
\begin{figure}[h]
    \centering
    \includegraphics[width=0.8\linewidth]{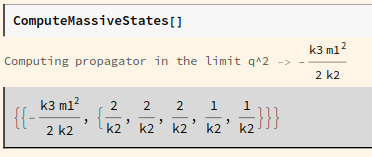}
    \caption{Masses and residues for the propagator of a symmetric propagation of massive spin-2.}
    \label{fig:TradMass2}
\end{figure}

Attempts to extend the Stueckelberg procedure to the case of Fierz-Pauli have resulted in models for spin-2 massive propagation reliant on the cooperation of multiple fields and intertwined gauge symmetries. By imposing invariance under a double gauge symmetry generated by the vector parameter $\xi_m(x)$ and $\Omega(x)$   
\begin{align} \label{gaugeTrad}
&     \delta^0 S(x) = \Omega(x)\, , \notag \\
&     \delta^0 V_m(x) =m_1 \xi_m(x) + \partial_m \Omega(x)\, , \notag \\
&     \delta^0 H_{mn}(x) = \partial_m \xi_n(x) + \partial_n \xi_m(x) \,   
\end{align}
the resulting model 
\begin{align}
\mathcal{L} = \;
&k_1 \Bigl[ H^{ab}\partial_b \partial_a H^c{}_c
- H^a{}_a\,\partial_c \partial_b H^{bc} \Bigr] \notag \\[6pt]
&+ k_2 \Bigl[ H^{ab}\partial_c \partial_b H_a{}^c
- \frac{1}{2} H^{ab}\Box H_{ab}
\notag \\[6pt]
& + \frac{1}{2} H^a{}_a\,\Box H^b{}_b
- H^a{}_a\,\partial_c \partial_b H^{bc} \Bigr] \notag \\[6pt]
&+ k_3 \Bigl[ \frac{m_1^2}{4} H_{ab} H^{ab}
- \frac{m_1^2}{4} H^a{}_a H^b{}_b
\notag \\[6pt]
&+ m_1 H^b{}_b\,\partial_a V^a
- m_1 H_{ab}\,\partial^b V^a \notag \\
&\phantom{+k_3\Bigl[}
+ m_1 H^{ab}\partial_b \partial_a S
- m_1 H^a{}_a\,\Box S \notag \\
&\phantom{+k_3\Bigl[}
+ \frac{1}{2} V^a \partial_b \partial_a V^b
- \frac{1}{2} V^a \Box V_a \Bigr]
\end{align}
is expected to propagate a massive spin-2 particle \cite{Hinterbichler:2011tt}. 
\texttt{Kummmitus} can easily take on this multicomponent model providing the unambiguous answer\footnote{The presence of a spurious massless pole is immediately recognized as such by a call to \texttt{ComputeMasslessStates[]} which gives an empty list.} Fig.~(\ref{fig:TradMass2}).
\subsection{Gauge-symmetric massive spin-2 - Part 2}
\begin{figure}[h]
    \centering
    \includegraphics[width=0.9\linewidth]{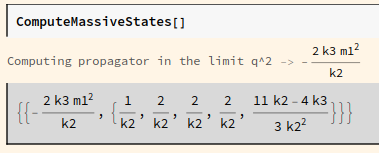}
    \caption{WTDIFF model for massive spin-2 propagation.}
    \label{fig:NonTrad}
\end{figure}

The double imposition of a five-parameter gauge shift on the four components of the vector $V_{m}$ in Eq.(\ref{gaugeTrad}) is a necessary, yet unpleasant\footnote{At the very least, it limits the gauge-fixing freedom in perturbative uses of the theory.}, feature of a Stueckelberg approach to the Fierz-Pauli mass combination of Eq.(\ref{FierzPauliMass}). It is reasonable to look at alternative gauge-invariant ways of generating a mass for a spin-2 particle. 
One among these (similar to an analogue one of \cite{Bonifacio:2015rea}) emerges using transversal diffeomorphism and a Weyl symmetry, with only an extra vector as a pure gauge field
\begin{align}
&     \delta^0 V_m(x) =m_1 \xi^T_m(x) + \partial_m \Omega(x)\, , \notag \\
&     \delta^0 H_{mn}(x) = \partial_m \xi^T_n(x) + \partial_n \xi^T_m(x) + m_1 \eta_{mn} \Omega(x)\,   .
\end{align}
The Lagrangian has the form 
\begin{align}
\mathcal{L} = \;
&k_1 \Bigl[ H^{ab}\partial_b \partial_a H^c{}_c
- H^a{}_a\,\partial_c \partial_b H^{bc} \Bigr] \notag \\[6pt]
&+ k_2 \Bigl[ H^{ab}\partial_c \partial_b H_a{}^c
- \frac{1}{2} H^{ab}\Box H_{ab}  \notag \\[6pt]
&
- \frac{1}{2} H^a{}_a\,\partial_c \partial_b H^{bc}
+ \frac{3}{16} H^a{}_a\,\Box H^b{}_b \Bigr] \notag \\[6pt]
&+ k_3 \Bigl[ m_1^2 H_{ab} H^{ab}
- \frac{m_1^2}{4} H^a{}_a H^b{}_b 
 \notag \\[6pt]
&+ m_1 H^b{}_b\,\partial_a V^a
- 4m_1 H_{ab}\,\partial^b V^a \notag \\
&\phantom{+k_3\Bigl[}
+ 2 V^a \partial_b \partial_a V^b
- 2 V^a \Box V_a  \notag \\[6pt]
&
+ H^a{}_a\,\partial_c \partial_b H^{bc}
- \frac{1}{4} H^a{}_a\,\Box H^b{}_b \Bigr] \, , 
\end{align}
and, via \texttt{Kummitus} FIG.~(\ref{fig:NonTrad}), 
we find the five propagating states and simple unitarity conditions over the parameter space. 
\subsection{Massless spin-3: Fronsdal}
The catalogue of projector operators up to rank-3, as computed in \cite{Mendonca:2019gco,Percacci:2020ddy} and completed for mixed field scenarios in \cite{Marzo:2021esg}, is part of an external library called by \texttt{Kummitus} (and which can be properly expanded to include higher-spin analyses\cite{Bittencourt:2025roa}). We can test the robustness of our method against spin-3 models starting with the pioneering realization given by Fronsdal \cite{Fronsdal:1978rb} 
\begin{align}
\mathcal{L} = \;
\frac{k_1}{2} \Bigl[
& \partial_l A_{ijk}\,\partial^l A^{ijk}
+ 3\,\partial_j A^l{}_{li}\,\partial^j A^k{}_{k}{}^i
- 6\,\partial_j A^l{}_{li}\,\partial_k A^{jki} \notag \\
&+ 3\,\partial_j A^j{}_{ik}\,\partial_l A^{lik}
+ \frac{3}{2}\,\partial^j A^l{}_{lj}\,\partial_i A^k{}_{k}{}^i
\Bigr] \, ,
\end{align}
obtained by imposing, to the generic quadratic Lagrangian, the symmetry\footnote{That all high-spin Lagrangians can be recovered demanding invariance under a reduced generalized diffeomorphisms has been recently proven in  \cite{Barker:2025bpe}.}
\begin{align}
&\delta A_{abc} = \partial_a \xi_{bc} + \partial_b \xi_{ac} + \partial_c \xi_{ab} \notag \\
& \xi_{ab} = \xi_{ba}\,, \qquad \xi^a{}_a = 0 \, . 
\end{align}

\begin{figure}[h]
    \centering
    \includegraphics[width=1.05\linewidth]{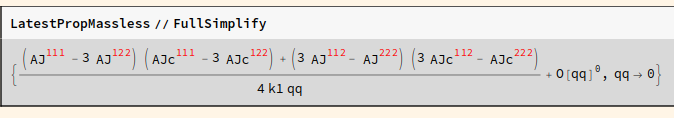}
    \includegraphics[width=1.05\linewidth]{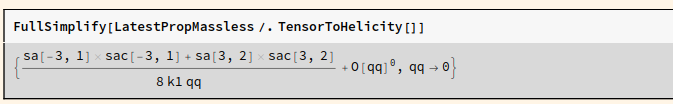}
    \caption{Saturated propagator for Fronsdal theory before and after transition to helicity-basis.}
    \label{fig:Fronsdal}
\end{figure}

%$$
Given that the rank-3 tensor \texttt{A[m,n,r]} is totally symmetric by default in \texttt{Kummitus}, we can proceed without further adjustments in calling \texttt{ComputeMasslessData} and immediately recognize the two propagating states and the following unitary request of $k_1>0$. Also for this case, a direct component-based access is largely (cosmetically) improved by a transition to the helicity basis FIG.(\ref{fig:Fronsdal}) 

\subsection{Spin-3 and Spin-1: Campoleoni-Francia}
\begin{figure}[h]
    \raggedright
    \includegraphics[width=0.6\linewidth]{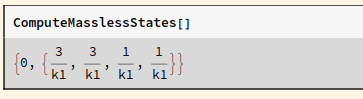}\\[4pt]
    \includegraphics[width=1.04\linewidth]{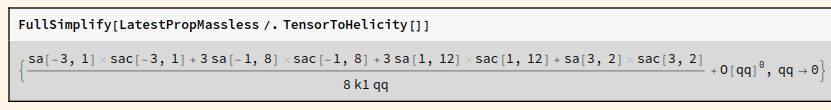}
    \caption{Multisector propagation in the Campoleoni-Francia model.}
    \label{fig:Francia}
\end{figure}

An even simpler theory exists which can accommodate a gauge-symmetric model for the propagation of spin-3. This was found and generalized for all spins in \cite{Campoleoni:2012th}. For spin-3, this admits the minimal (Maxwell-like) form
\begin{align}
\mathcal{L} = \;
\frac{k_1}{2} \Bigl[
A^{abc}\,\Box A_{abc}
- 3\, A^{abc}\,\partial_a \partial_s A^s{}_{bc}
\Bigr]\, , 
\end{align}
presenting four propagating states (and a simple request for unitarity $k_1 >0$) Fig.~(\ref{fig:Francia})
That such four states are the simultaneous propagation of helicity-3 and helicity-1 is visible from the helicity-basis expansion (also shown in Fig.~(\ref{fig:Francia})). 
\subsection{Massive spin-3, totally symmetric case: Singh-Hagen}
\begin{figure}[h]
    \includegraphics[width=0.7\linewidth]{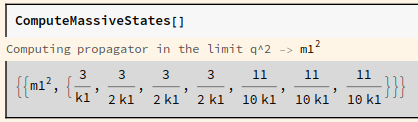}
    \caption{The seven propagating states of the Singh-Hagen model.}
    \label{fig:SinghHagen}
\end{figure}

Totally symmetric fields high-rank fields require support from lower-rank partners in order to accommodate for massive propagation. The Singh-Hagen model \cite{Singh:1974qz} is the simplest instance of this, demanding the presence of a rank-0 field $S(x)$:  
\begin{align}
\mathcal{L} = \;
&k_1 \Bigl[ m_1^2 A_{amn} A^{amn} 
- 3m_1^2 A^a{}_{am} A^{mn}{}_{n} \notag \\
&\phantom{+k_1\Bigl[}
+ A^{amn}\Box A_{amn}
- 3 A^a{}_{am}\partial^m \partial_r A^{nr}{}_{n}
\notag \\
&
- 3 A^{amn}\partial_n \partial_r A_a{}_m{}^r 
+ 6\, A^a{}_{am}\partial_r \partial_n A^{mn}{}^r
\notag \\
&
- 3\, A^a{}_{am}\Box A^{mn}{}_{n}
\Bigr] + k_3 \Bigl[
m_1 A^a{}_{am}\,\partial_a S
\Bigr] \notag \\[6pt]
&+ \frac{k_3^2}{k_1} \Bigl[
2m_1^2 S^2
+ \frac{1}{2} S\,\Box S
\Bigr] \, . 
\end{align}
The massive spin-3 seven propagating dofs can then easily be recovered from the residues of the propagator FIG.(\ref{fig:SinghHagen}). 
\subsection{Massive spin-3, totally symmetric case gauge-invariant: Klishevich-Zinoviev}
\begin{figure}[h]
    \includegraphics[width=0.7\linewidth]{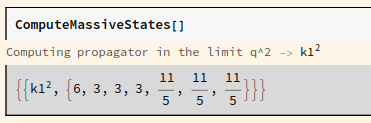}
    \caption{The seven propagating states of the Klishevich-Zinoviev model.}
    \label{fig:Zino}
\end{figure}

To propagate a massive spin-3 in a gauge-invariant model, even more auxiliary (pure-gauge) fields need to be invoked. The paradigmatic case has been provided by the analysis of \cite{Klishevich:1997pd,Zinoviev:2008ck} in an attempt to support consistent interactions with photons. The corresponding Lagrangian 
\begin{align}
\mathcal{L} = \;
&k_1^2 \Bigl[
\frac{1}{2} A_{abm} A^{abm}
- \frac{3}{2} A^a{}_{a}{}^b A_b{}^m{}_m
+ \frac{9}{4} H^a{}_a H^b{}_b
 \notag \\
& - 45\, H^a{}_a S + 225\, S^2
+ \frac{15}{2} A_a{}^b{}_b V^a
- \frac{45}{4} V_a V^a
\Bigr] \notag \\[6pt]
&+ k_1 \Bigl[
6\, A_b{}^m{}_m \partial_a H^{ab}
- \frac{3}{2} A_b{}^m{}_m \partial^b H^a{}_a
- 3\, A_{abm}\,\partial^m H^{ab} \notag \\
&\phantom{+k_1\Bigl[}
- 15\, H^b{}_b\,\partial_a V^a
+ 15\, H_{ab}\,\partial^b V^a
- 225\, V^a \partial_a S
\Bigr] \notag \\[6pt]
&+ \frac{3}{2} \partial_m H_{ab}\,\partial^m H^{ab}
- \frac{3}{2} \partial_b H^m{}_m\,\partial^b H^a{}_a
\notag \\[6pt]
&
- 3\,\partial_a H^{ab}\,\partial_m H_b{}^m
+ 3\,\partial^b H^a{}_a\,\partial_m H_b{}^m \notag \\[4pt]
&- \frac{1}{2}\,\partial_n A_{abm}\,\partial^n A^{abm}
+ \frac{3}{2}\,\partial_a A^{abm}\,\partial_n A_{bm}{}^n
\notag \\[6pt]
&
- 3\,\partial^m A^a{}_{a}{}^b\,\partial_n A_{bm}{}^n \notag \\
&+ \frac{3}{2}\,\partial_m A_b{}^n{}_n\,\partial^m A^a{}_{a}{}^b
+ \frac{3}{4}\,\partial_b A^a{}_{a}{}^b\,\partial_n A^m{}_{m}{}^n \notag \\[4pt]
&+ \frac{15}{2}\,\partial_a V_b\,\partial^b V^a
- \frac{15}{2}\,\partial_b V_a\,\partial^b V^a
+ 45\,\partial_a S\,\partial^a S \, ,
\end{align}
is a real test for the minimal, direct algorithm used by \texttt{Kummitus}, given the presence of many gauge constraints to account for. Nevertheless, after some minutes, the simple result can be accessed FIG.~(\ref{fig:Zino}), proving once again the successful accommodation of the wanted propagation.

\subsection{Massive spin-3 with hook symmetric field: Percacci-Sezgin}
\begin{figure}[h]
    \includegraphics[width=0.7\linewidth]{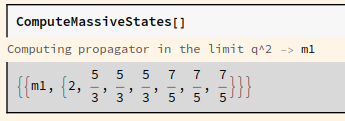}
    \caption{The seven propagating states of the Sezgin-Percacci model.}
    \label{fig:Percacci}
\end{figure}

Recently, new models for the propagation of a massive spin-3 modes without the need of auxiliary fields have been presented in \cite{Percacci:2025oxw}. The key, is in relying on 2-index symmetric tensor $A_{smn} = A_{snm}$ instead of a totally symmetric one. To prepare \texttt{Kummitus} for this case, we have to set the corresponding global variable to the correct value \texttt{rank3SymmetryType = "Sym23"} \emph{before} loading the package. All four models of \cite{Percacci:2025oxw} can be found in \texttt{Models.nb}. Here, for brevity, we focus on 'Model 1':
\begin{align}
\mathcal{L} = \;
& - m_1 \Bigl[
\frac{13}{12} A_{abc} A^{abc}
- \frac{19}{12} A^{ab}{}_{c} A_{ba}{}^{c}
- \frac{5}{2} A_a{}^b{}_b A^{ac}{}_c
\Bigr] \notag \\[6pt]
&- \frac{1}{6}\,\partial_d A_{abc}\,\partial^d A^{abc}
- \frac{1}{3}\,\partial_d A_{bac}\,\partial^d A^{abc} \notag \\[4pt]
&+ \frac{1}{6}\,\partial_a A^{abc}\,\partial_d A^d{}_{bc}
+ \frac{1}{3}\,\partial_b A^{abc}\,\partial_d A_{ac}{}^d
\notag \\[6pt]
&+ \frac{1}{3}\,\partial_b A^{abc}\,\partial_d A_{ca}{}^d
+ \frac{2}{3}\,\partial_a A^{abc}\,\partial_d A_{bc}{}^d \notag \\[4pt]
&- \frac{1}{3}\,\partial^c A^{ab}{}_b\,\partial_d A_{ac}{}^d
- \frac{2}{3}\,\partial^c A^a{}_{ab}\,\partial_d A_{bc}{}^d  
\notag \\[6pt]
&- \frac{2}{3}\,\partial^c A^{ab}{}_b\,\partial_d A_{ca}{}^d
- \frac{4}{3}\,\partial^c A^a{}_{ab}\,\partial_d A_{cb}{}^d \notag \\[4pt]
&- \frac{11}{18}\,\partial_a A^{ab}{}_b\,\partial_c A^{cd}{}_d
- \frac{13}{36}\,\partial_b A^a{}_{a}{}^b\,\partial_c A^{cd}{}_d
\notag \\[6pt]
&+ \frac{23}{144}\,\partial_b A^a{}_{a}{}^b\,\partial_d A^c{}_{c}{}^d + \frac{1}{6}\,\partial_c A_a{}^d{}_d\,\partial^c A^{ab}{}_b
\notag \\[6pt]
&+ \frac{2}{3}\,\partial_c A_{ab}{}^d{}_d\,\partial^c A^a{}_{a}{}^b
+ \frac{2}{3}\,\partial_c A^a{}_{ba}{}^d{}_d\,\partial^c A^b{}_{a}{}^a \, . 
\end{align}
Again, the successful propagation of a massive spin-3 state is confirmed FIG.~(\ref{fig:Percacci})

\section{Conclusions}
We have presented \texttt{Kummitus}, a minimal Wolfram Mathematica toolbox designed to compute the gauge-invariant saturated propagator and unambiguously identify its single-pole structure, providing direct access to the propagating degrees of freedom and their physical norm. The tool follows the shortest possible algorithmic path to the propagator, proceeding directly from the definitions of the spin-projector operator framework, with no detours or auxiliary constructions. To our knowledge, such a direct implementation has been absent from the publicly available literature, and \texttt{Kummitus} is intended to fill this role in a transparent and pedagogically accessible way.
The toolbox has been validated against a representative set of models, a selection of which has been presented here. A broader, growing collection, spanning from textbook cases to recent higher-spin models, is available in the accompanying notebook \texttt{Models.nb}, which is intended both as a test suite and as a practical reference for the user.
Used in conjunction with \texttt{PSALTer}~\cite{Barker:2024juc}, the two packages together provide a multi-pronged computational attack on the spectral problem. The availability of two independent tools, each approaching the problem from a different direction, offers a valuable cross-checking capability and broadens the range of models that can be efficiently analyzed.
\\
\begin{acknowledgments}
I am grateful to Will Barker, Dario Francia and Alessandro Santoni for precious discussions. 
I also thank the Department of Physics at the University of Calabria for its hospitality during my visit.
This work was supported by the Estonian Research Council grant PRG1677 and the CoE program TK202 'Fundamental Universe'. 
\end{acknowledgments}

\bibliography{bibl}

\end{document}